\documentclass[preprint,preprintnumbers,amsmath,amssymb]{revtex4}
\usepackage{graphicx}

\begin{document}
\title{Quintessential inflation from a
variable cosmological constant in a 5D vacuum}
\author{
$^{1}$Agustin Membiela\footnote{
E-mail address: membiela@argentina.com}
and $^{1,2}$Mauricio Bellini\footnote{
E-mail address: mbellini@mdp.edu.ar}}
\address{
$^1$Departamento de F\'{\i}sica,
Facultad de Ciencias Exactas y Naturales,
Universidad Nacional de Mar del Plata,
Funes 3350, (7600) Mar del Plata, Argentina.\\
$^2$ Consejo Nacional de Ciencia y Tecnolog\'{\i}a (CONICET).}

\vskip .5cm

\begin{abstract}
We explore an effective 4D cosmological model for the universe
where the variable cosmological constant governs its evolution and
the pressure remains negative along all the expansion. This model
is introduced from a 5D vacuum state where the (space-like) extra
coordinate is considered as noncompact. The expansion is produced
by the inflaton field, which is considered as nonminimally coupled
to gravity. We conclude from experimental data that the coupling of
the inflaton with gravity should be weak, but variable in
different epochs of the evolution of the universe.
\end{abstract}
\maketitle \vskip .2cm \noindent
Pacs numbers: 04.20.Jb, 11.10.kk, 98.80.Cq \\
\vskip 1cm
\section{Introduction}

Recent years have witnessed a large amount of interest in higher-dimensional
cosmologies where the extra dimensions are noncompact. A popular example
is the so-called Randall-Sundrum Brane World (BW) scenario\cite{1}.
Particular interest revolves around solutions which are not only Ricci
flat, but also Riemann flat ($R^A_{BCD}=0$), where the vanishing
of the Riemann tensor means that we are considering the analog of the
Minkowsky metric in 5D. An achievement of this theory is that all
the matter fields in 4D can arise from a higher-dimensional vacuum.
One starts with the vacuum Einstein field equations in 5D and
dimensional reduction of the Ricci tensor leads to an effective
4D energy-momentum tensor\cite{2}. For this reason the Space-Time-Matter
(STM) theory is also called Induced-Matter (IM) theory. BW and IM theories
may appear different, but their equivalence has
been recently shown by Ponce de Leon\cite{3}.

The potential energy of the scalar field and/or the presence of a
variable cosmological term could drive inflation, resolving puzzles
such as the monopole, horizon and flatness problems\cite{4}.
The variable cosmological term has also been mentioned as a possible
solution to the cosmological ``constant'' problem\cite{5} and, most
recently, as a candidate for the dark matter (or quintessence) making up
most of the Universe\cite{6}. A mechanism for obtaining the decay
of the cosmological parameter consists in relax $\Lambda$ to its
small present day value\cite{c1,c2,c3}.

In this letter we are aimed to study the evolution of the universe
which is governed by a variable cosmological constant
($\dot\Lambda <0$) in a 5D vacuum state, such that the expansion
of the universe is due to a scalar field (the inflaton field)
coupled to gravity. However, on an effective 4D metric the
universe evolves with an equation of state with negative pressure
${\rm p}$ (but with ${\rm p} \ge -\rho$). This kind of expansion
is the well known quintessential inflation\cite{QI}.

 To describe a scalar field $\varphi$, which is
nonminimally coupled to gravity in a 5D vacuum state, we consider
the metric\cite{7}
\begin{equation}\label{1}
dS^2 = \psi^2 \frac{\Lambda(t)}{3} dt^2 - \psi^2 e^{2\int
\sqrt{\frac{\Lambda}{3}} dt} dr^2 - d\psi^2,
\end{equation}
with the action
\begin{equation} \label{2}
I=
- {\Large\int} d^4x \  d\psi
\sqrt{\left|\frac{ {^{(5)}g}}{{^{(5)}g}_0}\right|}
\left[ \frac{{^{(5)}{\rm R}} }{16 \pi G} + \frac{1}{2} g^{AB} \varphi_{,A}
\varphi_{,B} - \frac{\xi}{2} \  ^{(5)} {\rm R} \varphi^2 \right].
\end{equation}
Here, $\Lambda(t)$ is the decaying cosmological constant ($\dot\Lambda <0$),
$G=M^{-2}_p$ is the
gravitational constant ($M_p=1.2 \times 10^{19} \  GeV$ is
the Planckian mass),
$\xi$ is the coupling of $\varphi$
with gravity and $^{(5)}{\rm R}$ is the Ricci scalar, which is zero
on the $R^A_{BCD}=0$ (flat) metric (\ref{1}). This metric is also
3D spatially isotropic, homogeneous and
flat. In such metric $dr^2 = dx^2+dy^2+dz^2$,
$\psi$ describes the fifth space-like coordinate and
$t$ is the cosmic time.
The Lagrange equations give us the equation of motion for $\varphi$
\begin{equation}\label{3}
\ddot\varphi + \left[ 3 \sqrt{\frac{\Lambda}{3}} - \frac{\dot\Lambda}{
\Lambda} \right] \dot\varphi - \frac{\Lambda}{3}
e^{-2 \int \sqrt{\frac{\Lambda}{3}} \  dt} \nabla^2_r \varphi
- \frac{\Lambda}{3}
\left[ 4 \psi \frac{\partial\varphi}{\partial\psi} + \psi^2
\frac{\partial^2\varphi}{\partial\psi^2} \right]
+ \xi \  ^{(5)} {\rm R} \varphi=0,
\end{equation}
such that the last term in (\ref{3}) is zero, because the metric
(\ref{1}) is flat.
Furthermore, the commutation expression between $\varphi$ and
$\Pi^t = {\partial {\cal L}\over \partial \varphi_{,t}} =
{3 \over \Lambda \psi^2} \dot\varphi$, is
\begin{equation}
\left[\varphi(t,\vec r,\psi), \Pi^t(t,\vec{r'},\psi')\right] =
\frac{i}{a^3_0} \  g^{tt} \left|\frac{^{(5)} g_0}{^{(5)} g}\right| \left(
\frac{\Lambda_0}{\Lambda}\right)
\delta^{(3)}(\vec r - \vec{r'}) \  \delta(\psi - \psi'),
\end{equation}
where $a_0$ is the scale factor of the universe when inflation starts.
The field $\varphi(t,\vec r,\psi)$ can be
transformed as
\begin{equation}  \label{tr}
\varphi(t,\vec r,\psi) = e^{-\frac{1}{2} \int \left[
3\left(\frac{\Lambda}{3}\right)^{1/2} - \frac{\dot\Lambda}{\Lambda}
\right] dt} \left(\frac{\psi_0}{\psi}\right)^2 \chi(t,\vec r,\psi),
\end{equation}
such that, due to the fact ${\partial\varphi \over \partial\psi} =
-{2\over \psi} \varphi$,  the equation (\ref{3}) holds
\begin{equation}\label{4}
\ddot\varphi + \left[ 3 \sqrt{\frac{\Lambda}{3}} - \frac{\dot\Lambda}{
\Lambda} \right] \dot\varphi - \frac{\Lambda}{3}
e^{-2 \int \sqrt{\frac{\Lambda}{3}} \  dt} \nabla^2_r \varphi
+ \left[ \frac{2\Lambda}{3}+ \xi \  ^{(5)} {\rm R} \right]\varphi=0.
\end{equation}
Using the transformation (\ref{tr}), we obtain the equation of
motion for the field $\chi$
\begin{eqnarray}
\ddot\chi & -& \frac{\Lambda}{3}
e^{-2\int \left(\frac{\Lambda}{3}\right)^{1/2} dt}
\nabla^2_r \chi
- \left\{\frac{1}{4} \left[
3\left(\frac{\Lambda}{3}\right)^{1/2} - \frac{\dot\Lambda}{\Lambda}
\right]^2 \right. \nonumber \\
&+ &  \left. \frac{1}{2} \left[\frac{\dot\Lambda}{2} \left(\frac{3}{
\Lambda}\right)^{1/2} - \left(\frac{\ddot\Lambda}{\Lambda} -
\left(\frac{\dot\Lambda}{\Lambda}\right)^2\right)\right] -
\left(\frac{2\Lambda}{3} + \xi \  ^{(5)} {\rm R} \right) \right\}
\chi =0.     \label{ch}
\end{eqnarray}
The field $\chi$ can be written as a Fourier
expansion
\begin{equation}\label{fou}
\chi(t,\vec r,\psi) = \frac{1}{(2\pi)^{3/2}} {\Large\int} d^3k_r
{\Large\int} dk_{\psi} \left[ a_{k_r k_{\psi}}
e^{i(\vec{k_r}.\vec{r} +{k_{\psi}}{\psi})}
\xi_{k_r k_{\psi}}(t,\psi)
+
a^{\dagger}_{k_r k_{\psi}}
e^{-i(\vec{k_r}.\vec{r} +{k_{\psi}} {\psi})}
\xi^*_{k_r k_{\psi}}(t,\psi)\right],
\end{equation}
such that
\begin{equation}\label{comu}
\left[ \chi(t,\vec{r},\psi),\dot\chi(t,\vec{r'},\psi)\right] =
\frac{i}{a^3_0} \  \delta^{(3)}(\vec{r}
-\vec{r'}) \  \delta({\psi} - {\psi'}),
\end{equation}
and $\xi_{k_r k_{\psi}}(t,\psi) = e^{-i {k_{\psi}} {\psi}}
\bar\xi_{k_r k_{\psi}}(t)$. The commutator (\ref{comu}) is satisfied
for $\left[a_{\vec{k_r}k_{\psi}}, a^{\dagger}_{\vec{k_r'}k_{\psi}'}\right]=
\delta^{(3)}(\vec{k_r} - \vec{k_r'}) \  \delta(\vec{k_{\psi}}-\vec{k_{\psi}'})$
and  $\left[a^{\dagger}_{\vec{k_r}k_{\psi}}, a^{\dagger}_{\vec{k_r'}k_{\psi}'}
\right]=
\left[a_{\vec{k_r}k_{\psi}}, a_{\vec{k_r'}k_{\psi}'}\right]=0$, if the
following condition holds:
\begin{equation}\label{cond}
\bar\xi_{k_r k_{\psi}}(t) \dot{\bar\xi}^*_{k_r k_{\psi}}(t) -
\bar\xi^*_{k_r k_{\psi}}(t) \dot{\bar\xi}_{k_r k_{\psi}}(t) = \frac{i}{a^3_0}.
\end{equation}
where $\bar{\xi}_{k_r k_{\psi}}(t)$ satisfies the following equation of
motion:
\begin{eqnarray}
\ddot{\bar\xi}_{k_r k_{\psi}} & + & \left\{ \frac{\Lambda}{3} k^2_r
e^{-2 \int \left(\frac{\Lambda}{3}\right)^{1/2} dt} -
\left[\frac{1}{4} \left[
3\left(\frac{\Lambda}{3}\right)^{1/2} - \frac{\dot\Lambda}{\Lambda}
\right]^2 \right.\right.    \nonumber \\
& + & \left.\left. \frac{1}{2} \left[\frac{\dot\Lambda}{2} \left(\frac{3}{
\Lambda}\right)^{1/2} - \left(\frac{\ddot\Lambda}{\Lambda} -
\left(\frac{\dot\Lambda}{\Lambda}\right)^2\right)\right] -
\left(\frac{2\Lambda}{3} + \xi \  ^{(5)} {\rm R} \right) \right]\right\}
\bar{\xi}_{k_r k_{\psi}} =0. \label{xi}
\end{eqnarray}
Hence, since $\xi_{k_{r} k_{\psi}}(t,\psi) = e^{-i k_{\psi}\psi}
\bar\xi_{k_{r} k_{\psi}}(t)$, the expansion
(\ref{fou}) now can be written as
\begin{equation}\label{fou1}
\chi(t,\vec r) = \frac{1}{(2\pi)^{3/2}} {\Large\int} d^3k_r
{\Large\int} dk_{\psi} \left[ a_{k_r k_{\psi}}
e^{i\vec{k_r}.\vec{r}}
\bar\xi_{k_r k_{\psi}}(t)
+ c.c. \right],
\end{equation}
being $c.c$ the complex conjugate.\\

\section{Effective 4D dynamics}

We consider the metric (\ref{1}). On the hypersurface
$\psi = \sqrt{{3 \over \Lambda(t)}}$, the effective 4D metric
that results is
\begin{equation} \label{5}
dS^2_{eff} = \left[1- \frac{3\dot\Lambda^2}{4\Lambda^3}\right] dt^2 -
\frac{3}{\Lambda} e^{2\int \sqrt{\frac{\Lambda}{3}} dt} dr^2,
\end{equation}
so that the effective 4D action for this metric is
\begin{equation}\label{6}
^{(4)} I = - {\Large\int} d^4 x \sqrt{\left|\frac{^{(4)} g}{^{(4)} g_0}
\right|} \left[\frac{ ^{(4)} {\rm R}}{16\pi G} + \frac{1}{2} g^{\mu\nu}
\varphi_{,\mu} \varphi_{,\nu} - \frac{\xi_{eff}}{2} ^{(4)} {\rm R} \varphi^2
\right],
\end{equation}
where $g_{\mu\nu} = {\rm diag}\left[ \left(1-{3\dot\Lambda^2\over
4\Lambda^3}\right), -a^2(t), -a^2(t), -a^2(t)\right]$, with
$a^2 (t)={3\over \Lambda}
e^{2\int \left(\frac{\Lambda}{3}\right)^{1/2} dt}$ and $\xi_{eff}$ is
the effective coupling of $\varphi$ with gravity
on the metric (\ref{5}).
Moreover,
$^{(4)} g$ is the determinant of $g_{\mu\nu}$ and $^{(4)} {\rm R}$ is the
Ricci scalar corresponding to the effective 4D metric (\ref{5}). For
this metric, we adopt the comoving frame given by the effective
4D velocities
\begin{equation}
u^{t} = \sqrt{\frac{4 \Lambda^3}{4\Lambda^3 -3 \dot\Lambda^2}},
\qquad u^r =0.
\end{equation}
The condition for the metric (\ref{5}) to be Lorentzian is
$g_{tt} >0$, which is valid when
\begin{equation} \label{cd}
4\Lambda^3 > 3 \dot\Lambda^2.
\end{equation}
Hence, in this letter we shall consider the case for which the condition
(\ref{cd}) complies along all the expansion of the universe.
The effective 4D Ricci scalar for the metric (\ref{5}) is
\begin{equation}\label{7}
^{(4)} {\rm R} = -2 \Lambda - \frac{24 \left[
2\sqrt{3} \Lambda^{11/2} \dot\Lambda
- 4 \Lambda^7 + 2 \Lambda^5 \ddot\Lambda
- \sqrt{3} \Lambda^{7/2} \dot\Lambda \ddot\Lambda \right]}{
\left[ 4 \Lambda^3 - 3 \dot\Lambda^2\right]^2}.
\end{equation}
From eqs. (\ref{5}) and (\ref{6}), we obtain
the equation of motion for $\varphi(t, \vec r) $
\begin{equation}\label{6'}
\ddot\varphi +
\left[\frac{\dot{\sqrt{|^{(4)} g|}}}{\sqrt{|^{(4)} g|}} +
\frac{\dot g^{tt}}{g^{tt}}
\right]
\dot\varphi - \frac{\Lambda}{3 g^{tt}}
e^{-2\int \sqrt{\frac{\Lambda}{3}} dt} \nabla^2_r \varphi +
 \xi_{eff} \frac{^{(4)} {\rm R}}{g^{tt}}  \varphi =0,
\end{equation}
where
\begin{eqnarray}
&& g^{tt} = \frac{4 \Lambda^3}{4\Lambda^3 - 3\dot\Lambda^2}, \\
&& \frac{\dot{\sqrt{|^{(4)} g|}}}{\sqrt{|^{(4)} g|}}+
\frac{\dot g^{tt}}{g^{tt}}
= 3 \left(\frac{\Lambda}{3}\right)^{1/2} +
\frac{3\left(\ddot\Lambda \dot\Lambda - 2\dot\Lambda \Lambda^2\right)}{
4\Lambda^3 - 3\dot\Lambda^2}.
\end{eqnarray}
Furthermore,
we can make the following identification in eq. (\ref{6'}):
\begin{equation}
V'(\varphi) =  \xi_{eff} \  \frac{^{(4)} {\rm R}}{g^{tt}} \varphi.
\end{equation}
From the action (\ref{6}), the effective 4D scalar potential
can be identified as
\begin{equation}
V(\varphi) =
\frac{\xi_{eff}}{2} \  ^{(4)} {\rm R} \varphi^2.
\end{equation}
Note that the effective 4D potential is due to the coupling of $\varphi$
with gravity. On the other hand the additional kinetic term
[${1\over 2}
g^{\psi\psi} \left(\varphi_{,\psi}\right)^2$] in the 5D
action (\ref{2}) has a dissipative effect in the 4D equation of motion
(\ref{6'}).
The commutation relation between $\varphi$ and $\Pi^t = {\partial
^{(4)} {\cal L} \over \partial \varphi_{,t}} =
g^{tt} \varphi_{,t}$ is
\begin{equation}\label{c1}
\left[ \varphi(t,\vec r), \Pi^t(t,\vec{r}')\right] =
\frac{i}{a^3_0} g^{tt} e^{-\int \left[
\frac{\dot{\sqrt{|^{(4)} g|}}}{\sqrt{|^{(4)} g|}}+
\frac{\dot g^{tt}}{g^{tt}}\right] dt} \delta(\vec r - \vec{r}').
\end{equation}
Using the transformation
\begin{equation}\label{tr1}
\varphi\left(t, \vec r\right)
=  \chi(t,\vec r)  e^{- \frac{1}{2} \int
\left[
\frac{\dot{\sqrt{|^{(4)} g|}}}{\sqrt{|^{(4)} g|}}+
\frac{\dot g^{tt}}{g^{tt}}
\right] dt},
\end{equation}
we obtain
\begin{equation} \label{chi1}
\ddot\chi -\left(\frac{4\Lambda^3 -3\dot\Lambda^2}{12 \Lambda^2}\right)
e^{-2\int \left(\frac{\Lambda}{3}\right)^{1/2} dt}
\nabla^2_r \chi - m^2(t) \chi =0,
\end{equation}
where
$m^2(t) = f^2(t) - \dot f(t) - \xi_{eff} g_{tt} ^{(4)} {\rm R}$, being
\begin{equation}
f(t) = -\frac{3}{2} \left[ \left(\frac{\Lambda}{3}\right)^{1/2} +
\frac{\dot\Lambda (\ddot\Lambda - 2\Lambda^2)}{(4\Lambda^3 - 3\dot\Lambda^2)}
\right].
\end{equation}
The field $\chi$ can be expanded in terms of their modes
$a_{k_r} e^{i \vec{k_r}.\vec{r}} \eta_{k_r}(t)$,
such that the equation of motion of the time-dependent modes
$\eta_{k_r}(t)$ are
\begin{equation}\label{eta}
\ddot{\eta}_{k_r} +
\left[ k^2_r \left(\frac{4\Lambda^3 - 3\dot\Lambda^2}{12 \Lambda^2}\right)
e^{-2\int \left(\frac{\Lambda}{3}\right)^{1/2} dt} - m^2(t) \right]
\eta_{k_r}(t)=0.
\end{equation}
Hence, from eq. (\ref{c1}) and eq. (\ref{tr1}), we obtain
\begin{equation}\label{c2}
\left[\chi(t,\vec r), \dot\chi(t,\vec{r}')\right] = \frac{i}{a^3_0}
\delta(\vec r - \vec{r}').
\end{equation}
The effective 4D equation of
state is ${\rm p} = \omega_{eff}(t) \rho$ (${\rm p}$ and
$\rho$ are respectively the pressure and the energy density), with
\begin{equation}\label{8}
\omega_{eff}(t) = -\frac{1}{3} \frac{\left[24 \Lambda^{9/2} + 18
\dot\Lambda^{3/2}-20\sqrt{3} \Lambda^3\dot\Lambda + 15 \sqrt{3} \dot\Lambda^3
-24 \Lambda^{5/2} \ddot\Lambda\right]}{
\left[ 8 \Lambda^{9/2} + 6 \dot\Lambda^2 \Lambda^{3/2} -
4\sqrt{3} \Lambda^3 \dot\Lambda + 3\sqrt{3} \dot\Lambda^3 - 2\sqrt{3}
\dot\Lambda^3 \Lambda\right]}.
\end{equation}
Note that when $\dot\Lambda =0$, one obtains
$^{(4)} {\rm R}=4\Lambda$ and the metric (\ref{5})
describes exactly an effective FRW metric with a pressure ${\rm p}_{v} =
- \rho_v = - {\Lambda \over 8\pi G}$, in a de Sitter expansion.

\section{An example}

We consider the case where $\Lambda(t) = 3 p^2(t)/t^2$, such
that $p(t)$ is given by
\begin{equation}\label{p(t)}
p(t) = 1.8 a t^{-n} + \left(\frac{b^2}{4a} - 0.95\right) + C \  t,
\end{equation}
where $a = {1\over 6} 10^{30 n} \  G^{n/2}$,
$b= {8\over 7} 10^{15 n} \  G^{n/4}$,
$C = 2 \times 10^{-61} \  G^{-1/2}$ and $n=0.352$.
There are at least four significative periods that we can identificate
in this model. \\

\begin{figure*}
\includegraphics[totalheight=8.5cm]{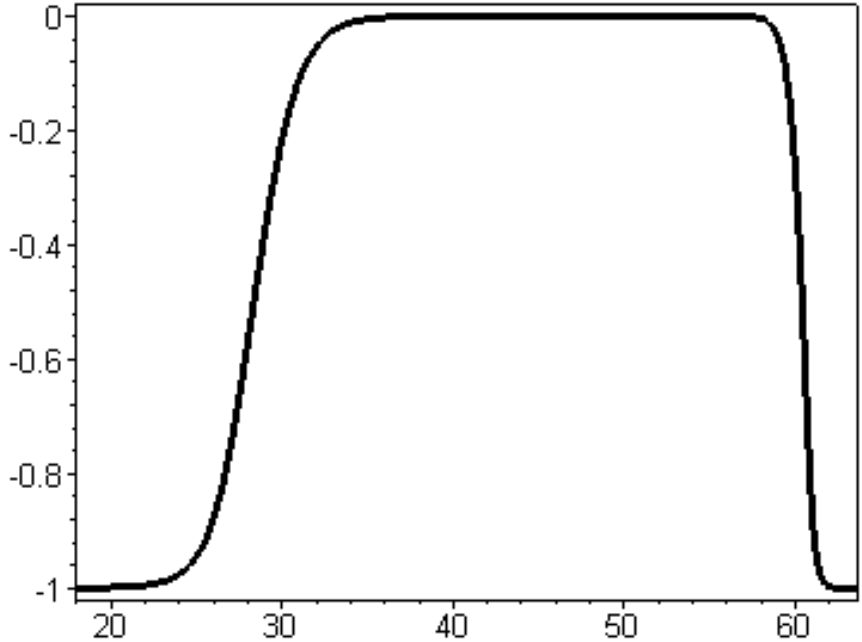} \caption{\label{f}  Evolution
of $\omega_{eff}[x(t)]$ as a function of $x(t) = {\rm
log}_{10}(t/t_p)$.\label{fig1}}
\end{figure*}

\noindent {\bf a)} The early period where the equation of state is
${\rm p} \simeq -\rho$, being $p(t) \gg 1$. In our model this
period holds for $t/t_p \ll 10^{10}$ [see fig. (\ref{f})]. In this
epoch we can make the approximation $\Lambda(t) \simeq
\Lambda_0$ (being $\Lambda_0$ a constant). \\

\noindent
{\bf b)} The period when
$p(t) \simeq 1+\epsilon_1$ ($\epsilon_1 = 0.000184$ in our model), where
the equation of state is nearly matter dominated (${\rm p} \simeq -
\epsilon_2 \rho$, being $\omega_{eff}=\epsilon_2 = -0.00025$ in our model).
In our model this epoch holds for $10^{35} < t/t_p < 10^{59}$ [see
fig. (\ref{f})].\\

\noindent
{\bf c)} The period which describes the present day universe, for
which $p(t) \simeq 1.898$ and the equation of state is ${\rm p} =
-0.687 \  \rho$. This epoch holds approximately when the universe
has an age $t/t_p \simeq 10^{60.652}$. \\

\noindent
{\bf d)} the asymptotic evolution of the
universe (for our model) where the expansion is a de Sitter expansion
with $p(t) = \Lambda_f$, being $\Lambda_f$ is the asymptotic final
value for the cosmological parameter. This epoch holds for
$t/t_p > 10^{62}$ [see fig. (\ref{f})]. \\

In the following subsections
we shall study with more detail
these different epochs for the evolution of the universe.

\subsection{Early (de Sitter)
inflationary period: $\Lambda \simeq \Lambda_0$}

The early inflationary period in which $p(t) \gg 1$, can be approximated
to a nearly de Sitter expansion where $\dot\Lambda^2/\Lambda^3 \ll 1$
and hence $\Lambda \simeq \Lambda_0 \simeq 3 p^2 /t^2_p$. In this
epoch, which describes the
expansion of the universe for $t \ll 10^{10} \  t_p$
(in our model), the general solution is given by
\begin{equation}
\eta_{k_r}(t) = A_1 {\cal H}^{(1)}_{\nu_1} [y_1(t)]
+ A_2 {\cal H}^{(2)}_{\nu_1}[y_1(t)],
\end{equation}
where $\nu_1=\sqrt{9-48 \xi_{eff}}/2$
and $y_1(t) = k_r e^{-\left({\Lambda_0\over 3}
\right)^{1/2} t}$.

The normalized solution with $A_1=0$ and $A_2= {i \over 2}
\sqrt{{\pi \Lambda_0 \over 3}}$, is
\begin{equation}
\eta_{k_r}(t) =
\frac{i}{2} \sqrt{\frac{\pi \Lambda_0}{3}}
{\cal H}^{(2)}_{\nu_1}\left[k_r e^{-\sqrt{\frac{\Lambda_0}{3}} t} \right].
\end{equation}
The power spectrum for the squared $\varphi$-fluctuations
calculated on scales $ k_r \gg e^{\sqrt{{\Lambda_0 \over 3}} t}$
(super Hubble scales), is
\begin{equation}\label{f1}
\left.\left<\varphi^2\right>\right|_{IR} \sim k^{3-2\nu_1}_r.
\end{equation}
It implies that this spectrum should be scale invariant only for
$\xi_{eff}=0$. In particular, we can calculate the range of validity for
$\xi_{eff}$ by comparing the spectrum (\ref{f1}) with observational
data\cite{PDG}, for the spectral index $n_s$
\begin{equation}\label{si}
n_s = 0.97 \pm 0.03.
\end{equation}
Making, $n_s -1 = 3-2\nu_1$, we obtain
\begin{equation} \label{r1}
 -0.008 < \xi_{eff} < 0.
 \end{equation}

\subsection{Matter dominated period: ${\rm p} = -\epsilon_2 \rho$}

In our model the equation of state which describes the expansion
of the universe for $10^{35} < t/t_p < 10^{59}$
[on the effective 4D metric (\ref{5})], is ${\rm
p}/\rho = -\epsilon_2 \simeq -0.00025 $, being $\Lambda(t) \simeq
{3 p^2/t^2}$ with $p \simeq (1+\epsilon_1) = 1.000184 $. In this
epoch, which describes an (asymptotic) matter dominated universe,
the general solution of the equation of motion (\ref{eta}) is
given by
\begin{equation}
\eta_{k_r}(t) = C_1 \sqrt{t} {\cal H}^{(1)}_{\nu_2}[y_2(t)] +
C_2 \sqrt{t} {\cal H}^{(2)}_{\nu_2}[y_2(t)],
\end{equation}
where $(C_1,C_2)$ are constants, ${\cal H}^{(1,2)}_{\nu_2}[y_2(t)]$ are
the first and second kind Hankel functions and $\nu_2 = {\sqrt{1+4
s(p)}\over 2p}$, $y_2(t) = k_r {\sqrt{p^2-1}\over p} \left({t\over
t_0}\right)^{-p}$. The normalized solution is
\begin{equation}
\eta_{k_r}(t) = \frac{i}{2} \sqrt{t} p \sqrt{\frac{\pi}{t^3_p}}
{\cal H}^{(2)}_{\nu_2}[y_2(t)],
\end{equation}
where
\begin{equation}
s(p) = \frac{3}{4} \left( 3p^2 + 4p +1\right)
-6 \xi_{eff} \left( 2 p^2 + 3p +1\right).
\end{equation}
The power spectrum for $\left<\varphi^2\right>$ on scales
$k_r \gg {p\over \sqrt{p^2-1}} \left(t/t_0\right)^p$, is
\begin{equation}\label{f2}
\left.\left< \varphi^2\right>\right|_{IR} \sim
k^{3-2\nu_2}_r,
\end{equation}
such that this spectrum is nearly scale invariant for $\xi_{eff}\simeq
0.111$. From experimental data (\ref{si}), we obtain
\begin{equation} \label{r2}
 0.109 < \xi_{eff} < 0.111.
 \end{equation}

\subsection{Present day epoch}

To study the present day epoch, which we estimate as $t_a \simeq
10^{60.652} \  t_p$, we can approximate $p(t)$ in eq. (\ref{p(t)})
\begin{equation}
p_a(t) \simeq \left(\frac{b^2}{4a} - 0.959\right) + C \  t.
\end{equation}
For $\left|{ t-t_a \over t_a}\right| \ll 1$, $p_a(t)$ can be
approximated to a constant, i.e.,
\begin{equation}
p_a(t) \simeq 1.898.
\end{equation}
Hence, this epoch can be treated as a power-law expanding
universe, with [see eq. (\ref{eta})]
\begin{equation}\label{m}
\ddot{\eta}_{k_r} + \left[ k^2_r \left(\frac{\Lambda_a(t)}{3 } -
\frac{ 0.16\Lambda_a(t)}{4} \right) e^{-2 C t_a} \left(\frac{t}{t_p}
\right)^{-2 p_a} - m^2_a(t) \right] \eta_{k_r} =0,
\end{equation}
where $\Lambda_a(t) = {3 p^2_a \over t^2}$,
\begin{equation}
m^2_a (t)= m^2(p=p_a,t) = \frac{9}{4} \frac{\left(p_a
+1\right)^2}{t^2} - \frac{3}{2} \frac{\left(p_a + 1\right)}{t^2} -
\frac{6 \xi_{eff}}{t^2} \left(p_a +1\right) \left(2 p_a + 1\right),
\end{equation}
and we have done the approximation $\left.{\dot\Lambda^2\over
\Lambda^3} \right|_{t=t_a,p=p_a} \simeq 0.16$. In this epoch
$\omega_{eff} \simeq -0.68$. The normalized solution of eq.
(\ref{m}), in this quintessential epoch, is
\begin{equation}
\left.\eta_{k_r}(t) \right|_{{|t-t_a| \over t_a}\ll 1} \simeq
\frac{i}{2} p_a \sqrt{\frac{\pi t}{t^3_a}} \  {\cal H}^{(2)}_{\nu_3} \left[
y_3 (t) \right],
\end{equation}
where $\nu_3 = {\sqrt{1+ 4B} \over 2 p_a}$, $y_3(t) = k_r
{\sqrt{3} \over p_a} \left(t/t_a\right)^{-p_a}$, and
\begin{displaymath}
B = \frac{9}{4} \left(p_a + 1\right)^2 - \frac{3}{2} \left(p_a +1\right)
- 6 \xi_{eff} \left(p_a + 1\right) \left( 2 p_a +1\right).
\end{displaymath}
One can calculate the power of the spectrum for
$\left<\varphi^2\right>$ on scales $k_r \gg {p_a\over \sqrt{3}}
\left(t/t_a\right)^{p_a} $, such that
\begin{equation}\label{f3}
\left.\left<\varphi^2\right>\right|_{IR} \sim
k^{3-2\nu_3}_r,
\end{equation}
which is nearly scale invariant (i.e., $\nu_3 \simeq 3/2$) for
$\xi_{eff} \simeq 0.08$. From observational data for $n_s$ (\ref{si}),
we obtain
\begin{equation} \label{r3}
 -0.006 < \xi_{eff} < 0.08.
 \end{equation}

\subsection{Asymptotic de Sitter expansion}

In our model the final asymptotic expansion of the universe can be
approximated to a nearly de Sitter expansion where
$\dot\Lambda^2/\Lambda^3 \ll 1$ such that $\Lambda \simeq
\Lambda_f \simeq 3 C^2$, where we are considering $C= 2 \times
10^{-61} \  G^{-1/2}$ in the power (\ref{p(t)}). In this epoch,
which describes in our model the expansion of the universe for $t
> 10^{62} \  t_p$, the general solution is
\begin{equation}
\eta_{k_r}(t) = B_1 {\cal H}^{(1)}_{\nu_4} [y_4(t)]
+ B_2 {\cal H}^{(2)}_{\nu_4}[y_4(t)],
\end{equation}
such that
$\nu_4=\sqrt{9 - 48 \xi_{eff}}/2 $ and $y_4(t) = k_r e^{-C t}$.

The normalized solution with $B_1=0$ and
$B_2= {i \over 2} \sqrt{\pi} C $, is
\begin{equation}
\eta_{k_r}(t) = \frac{i}{2} C
\sqrt{\pi} {\cal H}^{(2)}_{\nu_4}\left[k_r e^{-C t}
\right].
\end{equation}
The power of the spectrum for the squared $\varphi$-fluctuations
on scales $k_r \gg e^{C t}$, is
\begin{equation}\label{f4}
\left.\left<\varphi^2\right>\right|_{IR} \sim
k^{3-2\nu_4}_r,
\end{equation}
such that this spectrum become scale invariant for $\xi_{eff} =0$.
Finally, from (\ref{si}) we obtain
\begin{equation} \label{r4}
 -0.08 < \xi_{eff} < 0.
 \end{equation}
\vskip 5.cm

\section{Final Comments}

In this letter we have studied a model which describes all the expansion
of the universe governed by a decreasing cosmological constant from a
5D vacuum state. When we take a foliation on the fifth (space-like)
coordinate $\psi(t) = \sqrt{{\Lambda(t) \over 3}}$, the effective
4D dynamics describes an universe which has a 4D equation of state
${\rm p} = \omega_{eff} \rho$, with $\omega_{eff} <0$.
In this model, the expansion of the universe is due to the inflaton
field, which is considered as nonminimally coupled to gravity.
We have calculated the spectrum for the inflaton field fluctuations on
cosmological scales in four different epochs of its evolution.
\begin{itemize}
\item In the early inflationary expansion $\omega_{eff} \simeq -1$
and we obtain that the spectrum of
$\left.\left<\varphi^2\right>\right|_{IR} $ is nearly scale
invariant for $-0.08 < \xi_{eff} < 0$. \item In the matter dominated
epoch the universe is well described by $\omega_{eff} \simeq
-0.00025$.
If we split $\rho$ as $\rho=\rho^{(m)}+\rho^{(v)}+
\rho^{(r)}$ (being $\rho^{(m)}$, $\rho^{(v)}$
and $\rho^{(r)}$ the energy densities due respectively
to matter, vacuum
and radiation of the total energy density $\rho$), we
can differenciate two different stages. In the first one (after inflation
ends), $\rho^{(v)} >
\rho^{(r)} > \rho^{(m)}$, but in the second one
$\rho^{(v)} > \rho^{(m)} > \rho^{(r)}$.
However, the spectrum of
$\left.\left<\varphi^2\right>\right|_{IR} $ along all this
stage is scale invariant for
$0.109 < \xi_{eff} < 0.111$. \item The third epoch describes the present
day universe (which is considered as $1.5 \times 10^{10}$ years
old), with $\omega_{eff} \simeq -0.68$. For the spectrum of
$\left.\left<\varphi^2\right>\right|_{IR} $ to be nearly scale
invariant we obtain that the coupling must be $-0.06 < \xi_{eff} <
0.08$. \item The asymptotic universe is described by a
cosmological constant with $\Lambda \simeq \Lambda_f = 3 C^2$
(being $C=2\times 10^{-61} \ G^{-1/2}$). In this stage (valid for
$t \gg 1/C$), the universe evolves as in a 4D de Sitter expansion
with $\omega_{eff} \simeq -1$, such that
$\left.\left<\varphi^2\right>\right|_{IR} $ is nearly scale
invariant for $-0.08 < \xi_{eff} < 0$ (as in the early inflationary
expansion).
\end{itemize}

In view of these results, we conclude that $\xi_{eff}$ cannot be
constant along the evolution of the universe. However, $\xi_{eff}$
should be very weak. In particular, in the present day
quintessential epoch, the experimental data suggests that the
coupling should be nearly zero ($-0.06 < \xi_{eff} < 0.08$). On the
other hand, during the early and future inflationary expansions,
observation suggests that $\xi_{eff}$ should be negative, but during the
(asymptotic) matter dominated epoch the coupling should be
positive.

 \vskip .3cm \noindent
{\bf Acknowledgements}\\
\noindent
A. M. acknowledges UNMdP for financial support.
M.B. acknowledges CONICET and UNMdP (Argentina) for financial
support.\\

\end{document}